\documentclass[preprint,showpacs,amsmath,amssymb]{revtex4}

\usepackage{amsmath}
\usepackage{graphicx}
\usepackage{epsfig}

\newcommand{\be}{\begin{equation}}
\newcommand{\ee}{\end{equation}}
\newcommand{\bma}{\begin{displaymath}}
\newcommand{\ema}{\end{displaymath}}

\begin{document}

\title{Rotating quantum liquids crystallize}

\author{S.M. Reimann$^1$, M. Koskinen$^2$, Y. Yu$^1$,
  and M. Manninen$^2$}

\affiliation{\sl $^1$Mathematical Physics, Lund Institute of Technology,
SE-22100 Lund, Sweden}

\affiliation{\sl $^2$NanoScience Center, Department of Physics,
FIN-40014 University of Jyv\"askyl\"a, Finland}

\date{today}

\begin{abstract} 
Small crystallites form when finite quantal systems 
are set highly rotating. This crystallization is 
independent of the statistics of the particles, and occurs for both 
trapped bosons and fermions. 
The spin degree of freedom does not change the tendency for localization. 
In a highly rotating state, the strongly correlated bosonic and fermionic
systems approach to that of classical particles.
\end{abstract}

\maketitle


In molecules and solids, the atoms may organize to well ordered 
geometrical structures which largely determine
their physical and chemical properties.  
The single atoms {\it localize} in the sense that the position of one atom 
determines the positions of the other atoms.
The situation becomes different when atoms are weakly interacting,
like in the case of helium, or trapped, cold atom gases. 
At sufficiently low temperatures, the atoms -- which may be either 
bosons~\cite{anderson1995,andrews1996,cornell2002,ketterle2002} or
fermions~\cite{greiner2003,jochim2003} --
may form a delocalized quantum liquid where the 
identity of the individual atoms has disappeared.  
In the liquid state, the inter-atom correlation weakens fast with distance, 
i.e., a position of one atom does not 
any more fix the positions of distant atoms.
The electron gas is the classic example, where the transition from a 
liquid to a localized ground state may occur: Already in 1934, it was 
predicted by Wigner~\cite{wigner1934} that in the limit of low densities,  
the electron gas may have a crystalline phase. 
Though initially discussed for the bulk,  
the so-called Wigner crystal may also occur in {\it finite} fermion  
systems.  
A well-known example are quantum dots~\cite{chakraborty1999,reimann2002},  
small electron puddles in a semiconductor heterostructure, where  
electron localization has been studied extensively  
in the last decade~\cite{jauregui1993,egger1999,yannouleas1999,reimann2000,reusch2001}.
A strong magnetic field is known to enhance localization even in the 
high-density 
limit~\cite{maksym1990,ruan1995,maksym2000,maksym1996,vojs1997,manninen2001,jain2004,yannouleas2004},
confirming the prediction of 
classical ground state geometries of charged particles in
a two-dimensional harmonic confinement~\cite{bolton1993,bedanov1994}. 

Leaning on the analogy between strong magnetic fields and rotation 
of a quantum system, many analogies between the physics of the 
fractional quantum Hall effect in electron systems, and Bose-Einstein 
condensates at high angular momenta have been 
drawn~\cite{wilkin1998,wilkin2000,viefers2000,manninen2001,cooper2001,regnault2003,regnault2003b,chang2005}.
The similarity of many-body spectra and particle localization 
between quantum dots in magnetic fields and finite, 
rotating Bose condensates was pointed out 
by Manninen {\it et al.}~\cite{manninen2001} in 2001. 
The existence of these crystalline boson phases beyond the 
Gross-Pitaevskii mean field regime was later confirmed 
by Romanovsky {\it et al.}~\cite{romanovsky2004}, 
pointing at the analogy with the Tonks-Girardeau~\cite{tonks} transition 
of the Bose gas in one dimension~\cite{kinoshita2004}. 

In this report, we demonstrate that more generally, repulsively interacting 
quantum particles -- no matter whether they obey bosonic or fermionic 
statistics -- localize when brought to extreme rotation, 
forming a rotational analog to the above mentioned Wigner crystal at low
electron densities. The localization is nearly
independent of the interparticle interaction and the statistics of the 
particles, and also occurs if the spin degree of freedom is considered. 
In the latter case, the many-body states form a very 
narrow band, clearly separated from high-lying collective excitations. 
At low angular momenta (and for large particle numbers), 
the rotational spectrum correspondingly shows localization of 
quasi-particles which are identified as 
vortices~\cite{toreblad2004,manninen2005}.
\smallskip

\noindent
{\bf The Model.}~~~
Let us consider a number  $N$ of 
interacting particles confined in
a two-dimensional harmonic trap. 
In the spinless case, the particles may be, for example,  
bosonic atoms with only one hyperfine species, 
or polarized spin-1/2 fermions, say  electrons with each of them
in a spin-up state. Including the spin degree of freedom 
only increases the phase space.
The Hamiltonian is
\be
H=\sum _i ^N  \left( {p_i^2\over 2\mu } + {1\over 2} \mu 
\omega ^2 r_i ^2 \right ) +\sum_{i<j}^N v(\vert {\bf r}_i-{\bf r}_j\vert)
\label{hamiltonian}
\ee
where $N$ is the number of particles with mass $\mu$,
${\bf r}=(x,y)$ a two-dimensional position vector, $\omega_0$ the 
oscillation frequency of the confining potential, and
$v(r)$ the {\it repulsive} interparticle interaction.
We will mainly consider long-range Coulomb interactions,
$v(r)={e^2}/{4\pi\epsilon_o r}~,$
but also demonstrate that a short range Gaussian interaction
$v(r)=v_0\exp(-r^2/2\sigma^2)$ results in similar localization of particles 
when the system is set rotating.
(Note that we do not explicitly include 
magnetic fields, which in absence of the Zeeman effect
only induce the rotation).

In the harmonic oscillator, the single-particle 
eigenstates are labeled by the number of radial 
nodes $n$, and by the angular momentum $l$ and its value 
with respect to the axis of quantization, $m$~. 
For the lowest-energy states at a given angular 
momentum $L$, the so-called {\it yrast } 
states~\cite{mottelson1999}, maximum alignment $L=M$ 
restricts the space to states with $n=0$ and $m\ge 0$. 
This is equivalent to the so-called 
lowest Landau level approximation frequently made for electron systems 
in strong magnetic fields.

The non-interacting, single-particle part of the many-body Hamiltonian 
contributes an excitation energy $(L+1)\hbar \omega $ to the ground state. 
In the lowest Landau level, it is thus sufficient to diagonalize only 
the interaction part of the Hamiltonian, $V_{int}({\bf r}_i, {\bf r}_j)$.
Consequently, the results are independent of the strength of the 
confinement (apart from the energy scale).
One can obtain the lowest-energy states and 
the energetically low-lying excitations at a given 
large value of angular momentum $L$ with high numerical accuracy. 
Naturally, for repulsive interactions, the  
interaction energy, $\langle V_{int}\rangle $, is lowered with increasing
angular momentum. 

\begin{figure}[h]
\includegraphics[width=\columnwidth]{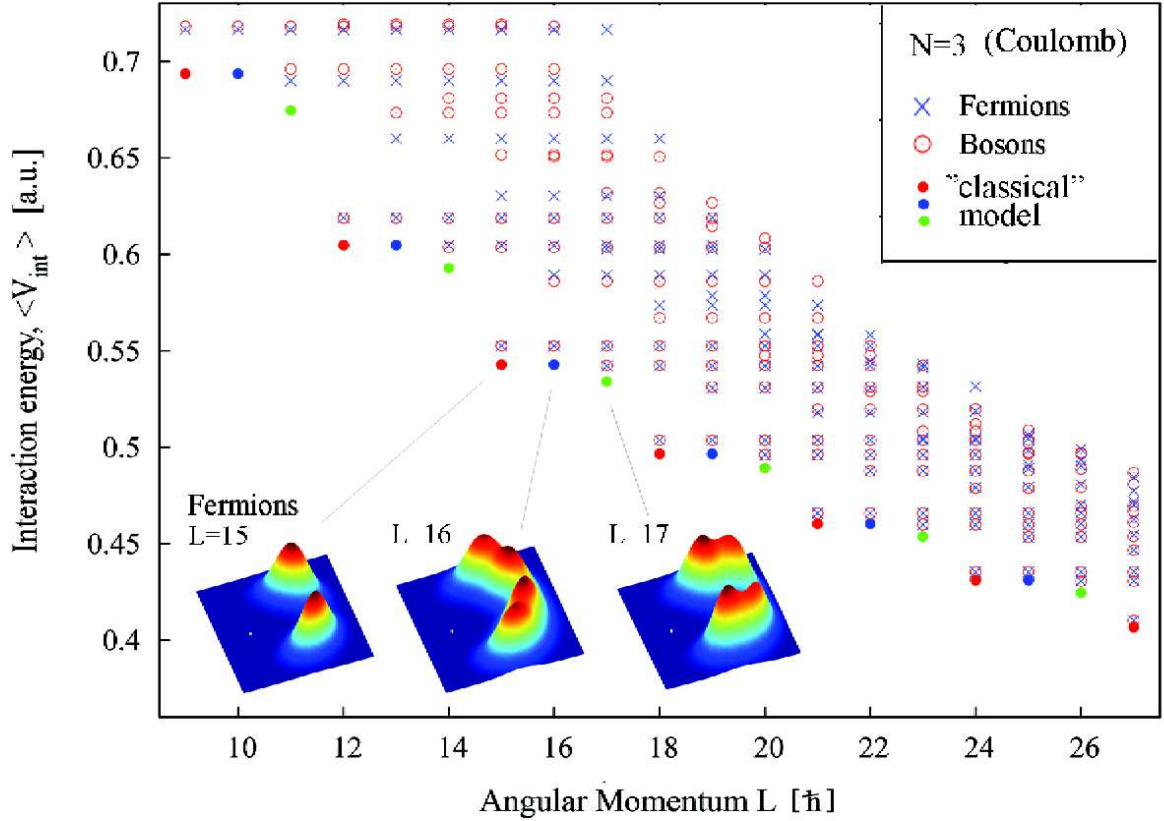}
\caption{Many-particle spectrum for three fermions (blue crosses) and
three bosons (red circles).
The red, blue and green dots show the result obtained by quantization
of the classical vibrations and rotations of the localized particles.
(Note, that only the interaction part of the energy is shown.)}
\label{three}
\end{figure}

\smallskip

\noindent
{\bf Spinless Wigner molecules.}~~~
Figure \ref{three} shows the many-particle spectra calculated for three particles
with Coulomb interaction. The results obtained for
bosons (open circles) approach those obtained for fermions (crosses) when the 
angular momentum increases. (The dots mark the energies  obtained by 
quantizing the {\it classical} rotations and vibrations, as explained below.)
The yrast line of the three-particle system 
shows pronounced cusps with period of three units of angular momentum. 
Such cusp states were discussed earlier in the context of quantum
dots at high magnetic fields~\cite{maksym1990,ruan1995,maksym2000,maksym1996,vojs1997,manninen2001,yannouleas2004}. In the above example, they originate 
from the rigid rotation of a triangular, molecule-like like structure of the 
three-body system. The symmetry group of the molecule dictates that a 
purely rotational state can be obtained at only every 
third angular momentum~\cite{ruan1995}. 

The localization of particles can be confirmed by looking at the
interparticle correlations. The insets to 
Fig. \ref{three} show the pair correlation functions 
for three fermions at angular momenta $L=15, 16 $ and 17.  
Clearly, at the cusps in the yrast line, the pair correlation
function resembles three localized electrons. In the case of
center-of-mass or vibrational excitation, the
peaks of the pair correlation function are smeared out due to the 
internal motion of the particles.

Let us now analyze the low-energy states 
in terms of the quantized rotational-vibrational 
spectrum. When the classical Wigner molecule
is set rotating, it expands and the eigenfrequencies of
vibrations change. 
In the case of a rigid rotation the classical energy 
is obtained by minimizing
\be 
E_{\rm cl}(L)=\frac{1}{2}\mu\omega_0\sum_i^N r_i^2
+\sum_{i<j}\frac{e^2}{4\pi\epsilon_0\vert{\bf r}_i-{\bf r}_j\vert}
+\frac{L^2}{2\mu \sum r_i^2}
\ee
with respect to the particle coordinates ${\bf r}_i$. 
However, in determining the classical vibrational
frequencies the Coriolis force has to be included.
We have chosen to solve the classical vibration frequencies using
molecular dynamics. The classical system of particles was
first set to rigid rotation, then small random changes to the particle
velocities were induced, and finally the eigenfrequencies were determined
from Fourier transforms of the interparticle distances.
It turned out that for particles with repulsive interactions confined 
in a harmonic well, the correction due to the Coriolis force is
essential as opposite to normal rotating molecules where it
only gives a small correction~\cite{landau1958}. 
For example, in the case of three classical electrons
in a harmonic trap,
the vibrational frequencies of a non-rotating Wigner molecule
are two-fold degenerate, with  $0.61/\omega_0$ and $0.87\omega_0$.
However, at high angular momenta, one frequency approaches zero 
while two frequencies approach $2\omega_0$.
(Note that irrespective of the angular momentum a center
of mass vibration has always the frequency $\omega_0$).  

Figure~\ref{three} displays the energies of the purely rotational states
as red points. These are followed by a state including
center of mass rotation (blue points), and a state where the rigid
rotation is accompanied by an internal vibration (green points).
The minor difference between the fully quantum-mechanical
spectrum and the quantized classical spectrum is caused by 
neglecting the zero-point energy of the low-energy vibrational mode.
In fact, also the low-energy excitations can be explained 
by multiples of center-of-mass and vibrational excitations of the 
quasi-classical molecule.
Only at high energies or low angular momenta, where the bosonic 
spectrum begins to deviate from the fermionic one, this 
simple picture looses its accuracy. 

The  spectrum in Figure~\ref{three} clearly 
shows that the low-energy states 
of three interacting particles confined in a two-dimensional harmonic
trap approaches that of three localized particles when the 
angular momentum of the system increases. Remarkably, the low-lying 
states at high angular momenta appear identical for bosons and fermions. 

\begin{figure}[h]
\includegraphics[width=0.9\columnwidth]{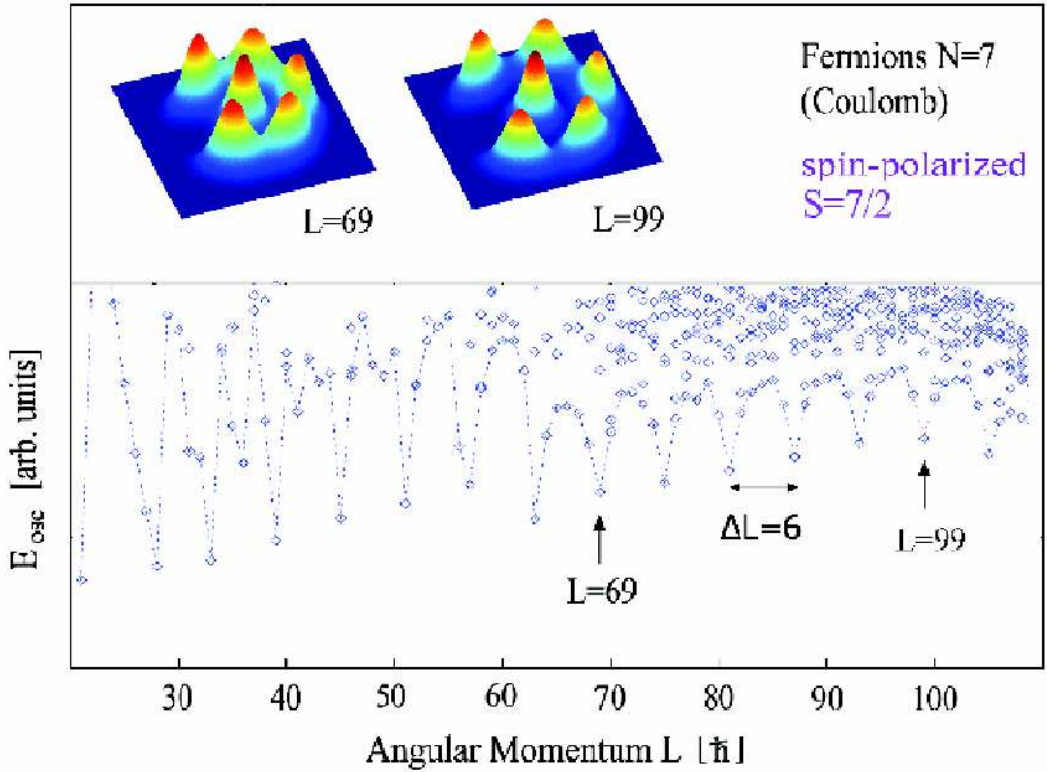}
\caption{Energy spectrum of seven polarized
electrons in a two-dimensional 
harmonic trap as a function of the angular momentum.
A smooth polynomial is subtracted from the potential to 
emphasize the oscillations caused by particle localization.
The inset shows examples of the pair correlation function. 
}
\label{sevenf}
\end{figure}

\begin{figure}[h]
\includegraphics[width=0.9\columnwidth]{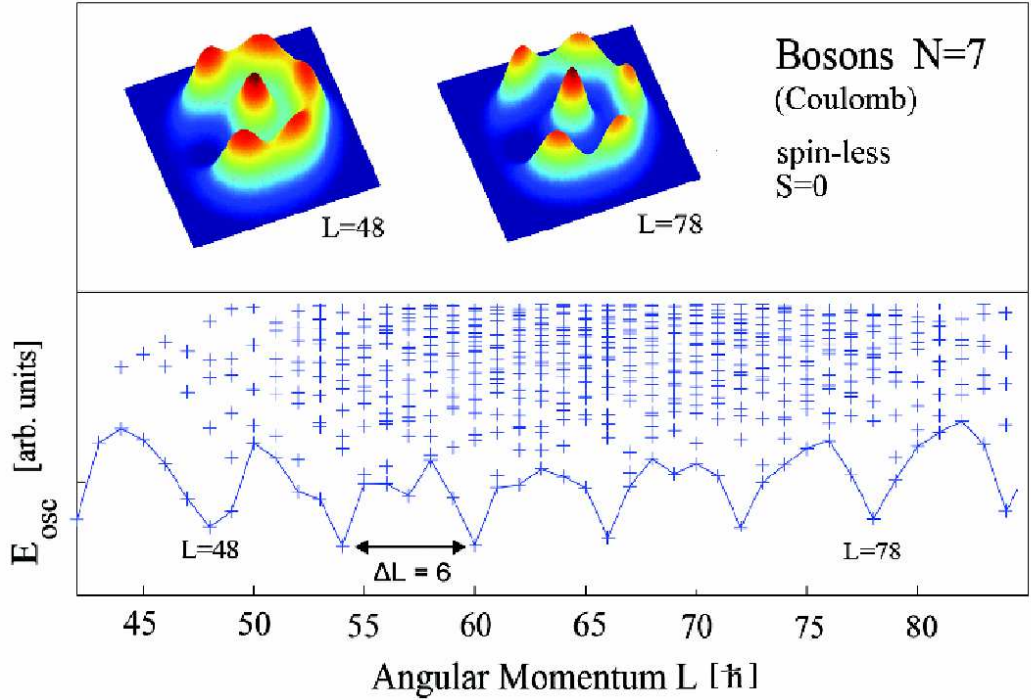}
\caption{Energy spectrum of seven bosons in a two-dimensional 
harmonic trap as a function of the angular momentum.
A smooth polynomial is subtracted from the potential to 
emphasize the oscillations caused by particle localization.
The inset shows examples of the pair correlation function.
The particles interact with the Coulomb interaction. 
}
\label{sevenb}
\end{figure}

To study whether the localization occurs also for larger particle
numbers we studied the many-particle spectra of fermions and bosons 
up to 7 particles. In all cases the spectrum directly reveals the
localization and its approach to the classical Wigner crystal energy
when the angular momentum increases. Figures~\ref{sevenf} 
and~\ref{sevenb} show the oscillations of the the interaction energy as a
function of angular momentum, comparing fermions with bosons. 
In both cases we subtracted a smooth second-order polynomial
from the spectra to illustrate the
characteristic oscillation of the yrast line as a function of angular
momentum. In this example with $N=7$, we observe a clear period of six. 
This is caused by the six-fold symmetry of the classical 
configuration~\cite{bolton1993,bedanov1994}, with one electron at the 
center and the remaining six electrons forming a hexagon around it.
The pair correlations shown in the inset confirm this classical geometry. 
They are displayed for a few points where the energy curve has a 
downward cusp.  The somewhat deeper minima at
 angular momenta $L=63$ and $L=105$ 
correspond to the fractional quantum Hall liquids with 
fractions (filling factors) 1/3 and 1/5.

Note that even if the oscillation period is the same for bosons and
fermions, there is a phase shift of three angular momenta between
these two results. This is caused by the different symmetry
requirements of the fermion and boson states~\cite{tinkham1964}.

\begin{figure}[h]
\includegraphics[width=0.9\columnwidth]{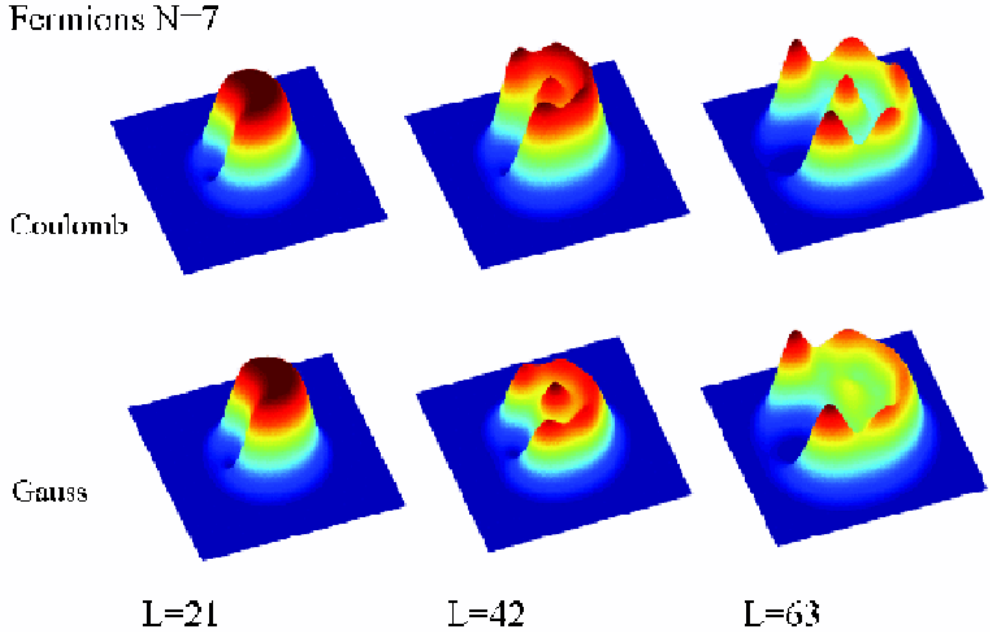}
\caption{Pair correlation functions for seven fermions interacting
with Coulomb interaction (upper row) and with short range Gaussian
interaction with $\sigma=0.05$ (lower row). The angular momenta
from left to right are $L=21$, 42, and 63.}
\label{cg}
\end{figure}

To illustrate that the localization indeed is caused by the increase
of angular momentum, we show in 
Fig. \ref{cg} the evolution of the pair correlation towards 
the localized state, beginning with small angular momenta. 
Two sets of results are shown, one for
the Coulomb interaction and one for a short range Gaussian
interaction. 
In both cases, the pair correlation function of the lowest
angular momentum shows no localization, but a smooth particle
distribution (corresponding to the 
so-called maximum density droplet~\cite{macdonald1993} as the 
finite-size analog to a quantum Hall liquid at integer filling factor).
When the angular momentum increases, the system expands
(due to the centrifugal force) and the localization gradually 
increases towards a Wigner molecule, as also shown in Figs. \ref{sevenf} and 
\ref{sevenb} above. Note, however, that the localization is 
weaker in the case of short-range interactions. 
The result shown in Fig. \ref{cg} is for fermions, but 
similar localization and expansion of the cluster of particles is
obtained in the case of bosons~\cite{romanovsky2004}. 

\smallskip
\begin{figure}[h]
\includegraphics[width=\columnwidth]{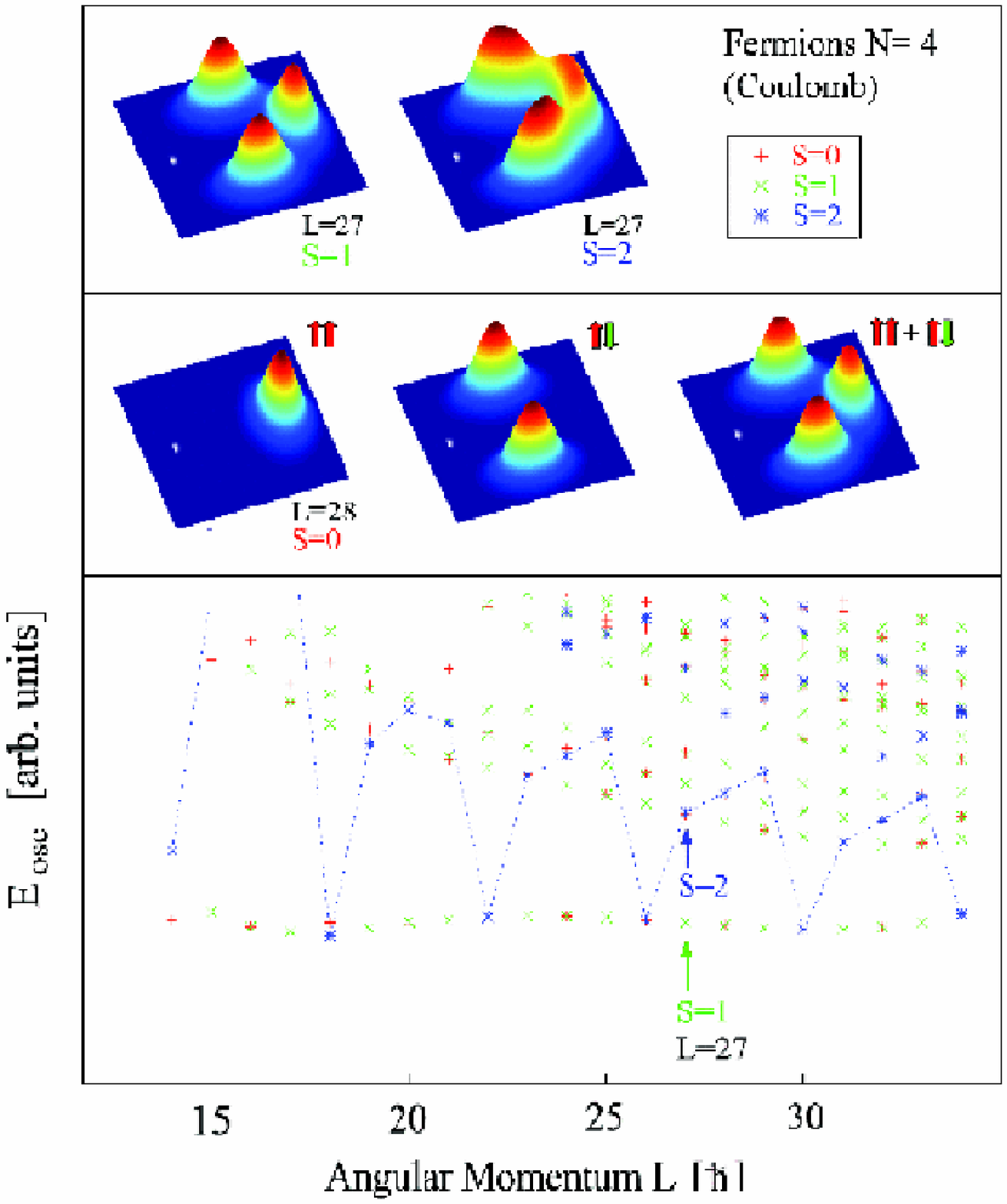}
\caption{Energy spectrum (lowest panel) of four electrons with spin as a 
function of the angular momentum. The different symbols correspond
to different total spin of the system.
The upper panel shows the total pair correlation function
for the two lowest states at $L=27$. The center panel 
shows the $\uparrow\uparrow $, $\uparrow-\downarrow $, and total 
($\uparrow+\downarrow$) pair correlations
for the $L=28$, $S=0$ state, indicating the anti-ferromagnetic
coupling of the localized electrons.}
\label{spin4}
\end{figure}

\noindent
{\bf Effect of spin.}~~~ 
An important question now is whether the spin degree of freedom 
neglected so far, may 
change the tendency of localization of fermions.
Notice that the Hamiltonian (\ref{hamiltonian}) is spin-independent and 
the only effect of spin is to increase the phase
space. 

Figure~\ref{spin4} shows the energy spectrum for four electrons
with spin. The different total spin states are shown with
different colors. The fully polarized states (blue asterisks connected by 
a dashed line) show now a period of four, consistent with a 
classical geometry where the electrons localize at the corners of a square.
When all spin states are allowed, the {\it lowest energy states 
form a narrow band, clearly separated from the higher excitations.}
We argue that these states are a consequence of  
rigid rotations of the Wigner molecule:
Now, as spin is included,  pure rotations are allowed for any angular momentum
due to the spin excitations, which cost nearly zero energy.

The situation is related to that in narrow quantum 
rings~\cite{koskinen2001,viefers2004}.
In strictly one-dimensional rings,
the spin and charge degrees of freedom completely separate
and the Heisenberg model, together with the quantized rotational-vibrational
states, can explain in quantitative detail the whole many-particle
spectrum~\cite{koskinen2001}.
In the case of a quantum dot, the situation seems to be
slightly more complicated: While the Heisenberg model
(and group theoretical analysis) still gives 
the correct spin states for each angular momentum, it does not 
explain quantitatively the energy differences, i.e.
the spin and charge excitations seem not to be completely decoupled.

Figure~\ref{spin4} also shows examples of correlation functions.

The upper panel shows the total correlation
functions  for  $L=27$ for the lowest state $S=1$ and the first excited
state $S=2$. Clearly the state $S=1$, belonging to the lowest band,
shows localization of the electrons. In the excited state, however, the
peaks are smeared out due to a center of mass excitation.
The middle panel shows the correlation function for the lowest
energy state with $L=28$. In this case the total spin is $S=0$ and
the state clearly shows anti-ferromagnetic order and electron
localization. 

\begin{figure}[h]
\includegraphics[width=0.9\columnwidth]{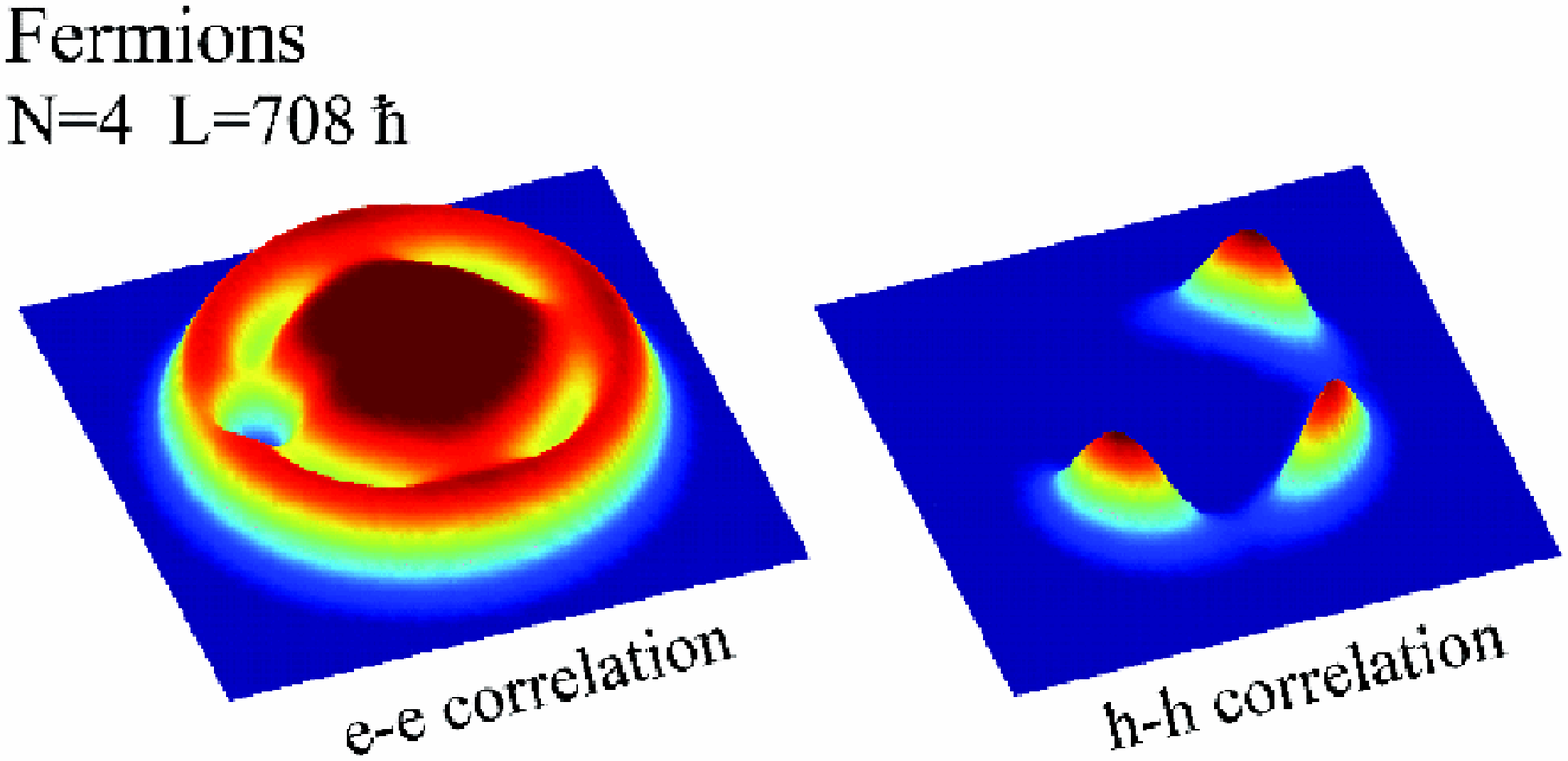}
\caption{Pair correlation function of 36 electrons in a harmonic
trap with angular momentum $L=708$ showing the exchange-correlation
hole and four vortices (left) and the vortex-vortex correlation
function showing that the vortices are well localized (right).
}
\label{vortices}
\end{figure}

When the number of electrons increases, 
computations including spin become more demanding, and it is
not possible to reach as high angular momenta as in the spectra
of Figs.~\ref{sevenf} and \ref{sevenb}. However, traces of
localization can be seen already at relatively low angular 
momenta, as indicated in Fig. \ref{cg}. 
We have made further calculations with up to seven particles. In all
cases,  the results showed a clear separation of the energy 
band, pointing at electron localization with possible spin-excitations,
similar to the four-electron case shown in Fig. \ref{spin4}.

\smallskip
\noindent
{\bf Localization of Vortices.}~~~
Finally, we note that localization by rotation
can also happen in the case of quasi-particles, even when they 
are collective excitations. 
An example here are the vortex patterns emerging in rotating clouds of
bosons or fermions at smaller angular 
momenta~\cite{toreblad2004,manninen2005,saarikoski2004}. 
For clear vortex states, naturally, the number of particles 
should be much larger than that of the vortex quasi-particles. 
We are thus limited to study polarized fermions (or spinless bosons). 
Figure \ref{vortices} shows the electron-electron 
pair-correlation function 
computed for 36 electrons with angular momentum $L=708$.
It should be emphasized that the angular momentum $L=708$, 
although a large number, is relatively {\it small}.
In fact, it corresponds to $L=24$ for seven particles,
where according to Fig. \ref{cg} no localization of particles
is expected.  

In addition to the exchange-correlation hole around the 
reference electron, we see four {\it minima} in the otherwise
smooth density distribution. 

The formation and localization of vortices can be 
understood with help of the electron-hole dualism~\cite{manninen2005},
which also makes it possible to plot the vortex-vortex correlation function 
(also shown in Fig. \ref{cg}).
Similar vortex localization was also found in bosonic 
systems~\cite{abo-shaer2001,butts1999,kavoulakis2002,toreblad2004}.

\smallskip

\noindent
{\bf Conclusions}~~~
We have shown that quantum mechanical particles,
when set rotating in a two-dimensional harmonic trap,
tend to localize to Wigner molecules. 
The localization is seen clearly in the periodic 
oscillations in energy spectrum as a function
of the angular momentum.
The many-particle spectrum can be explained in detail by
rigid rotation and vibrational modes calculated using 
classical mechanics. 
These results seem universal.  They are independent of the
shape or range of the interparticle interaction, and
fermions and bosons show similar localization. 
Considering also the spin degree of freedom
does not change the tendency for localization. 
With spin, the many-body states form a very 
narrow band, clearly separated from high-lying collective excitations. 
When the particle number is sufficiently large, the 
rotational spectrum shows localization of quasi-particles
which can be identified as vortices.


{\bf Acknowledgments}

We would like to thank B. Mottelson, J. Jain, S. Viefers and C. Yannouleas 
for several valuable discussions.

\end{document}